\documentclass[final,5p,times]{elsarticle}

\usepackage{amsmath}
\usepackage{orcidlink}
\usepackage{appendix}

\makeatletter
\renewcommand\appendix{\par
  \setcounter{section}{0}%
  \setcounter{subsection}{0}%
  \setcounter{equation}{0}%
  \setcounter{table}{0}
  \setcounter{figure}{0}
  \gdef\theequation{\@Alph\c@section.\arabic{equation}}%
  \gdef\thefigure{\@Alph\c@section.\arabic{figure}}%
  \gdef\thetable{\@Alph\c@section.\arabic{table}}%
  \gdef\thesection{\appendixname\@Alph\c@section}%
  \@addtoreset{equation}{section}%
  \@addtoreset{table}{section}
  \@addtoreset{figure}{section}
}
\makeatother

\journal{Physics Letters B}

\begin{document}

\begin{frontmatter}

\title{Probing hard/soft factorization via beam-spin asymmetry in exclusive pion electroproduction from the proton}

\author[REG]{A.C.~Postuma\orcidlink{0009-0008-4809-1251}}
\author[REG]{G.M.~Huber\orcidlink{0000-0002-5658-1065}}
\author[JLAB]{D.J.~Gaskell\orcidlink{0000-0001-5463-4867}}
\author[REG]{N.~Heinrich\orcidlink{0009-0005-8720-9329}}
\author[CUA,JLAB]{T.~Horn\orcidlink{0000-0003-1925-9541}}
\author[REG]{M.~Junaid\orcidlink{0000-0002-5853-4326}}
\author[REG,YORK]{S.J.D.~Kay\orcidlink{0000-0002-8855-3034}}
\author[REG]{V.~Kumar\orcidlink{0000-0003-0966-7365}}
\author[FIU]{P.~Markowitz\orcidlink{0000-0002-5661-586X}}
\author[OHIO]{J.~Roche\orcidlink{0000-0003-1913-3308}}
\author[CUA]{R.~Trotta\orcidlink{0000-0002-8193-6139}}
\author[REG]{A.~Usman\orcidlink{0000-0002-4344-7764}}
\author[KAU]{B.-G.~Yu\orcidlink{0000-0002-1075-6042}}
\author[YONSEI]{T.K.~Choi\orcidlink{0009-0006-9743-9088}}
\author[KAU]{K.-J.~Kong\orcidlink{0000-0002-0324-5690}}
\author[CUA]{S.~Ali\orcidlink{0000-0002-5596-1154}}
\author[REG]{R.~Ambrose\orcidlink{0009-0009-4189-8690}}
\author[ZAG]{D.~Androic\orcidlink{0000-0002-3921-5696}}
\author[TEMP,ANL]{W.~Armstrong\orcidlink{0000-0001-7359-6341}}
\author[WM]{A.~Bandari\orcidlink{0009-0006-5806-4164}}
\author[CUA]{V.~Berdnikov\orcidlink{0000-0003-1603-4320}}
\author[MSU]{H.~Bhatt\orcidlink{0000-0003-0087-5387}}
\author[MSU]{D. Bhetuwal\orcidlink{0000-0002-1006-9744}}
\author[HU]{D.~Biswas\orcidlink{0009-0001-1427-979X}}
\author[TEMP]{M.~Boer\orcidlink{0000-0002-4229-3935}}
\author[WM]{P.~Bosted\orcidlink{0009-0002-5543-0991}}
\author[CNU]{E.~Brash\orcidlink{0000-0002-3480-0394}}
\author[JLAB]{A.~Camsonne\orcidlink{0000-0003-4333-2614}}
\author[JLAB]{J.P.~Chen\orcidlink{0000-0002-1428-1646}}
\author[WM]{J.~Chen}
\author[UVA]{M.~Chen\orcidlink{0000-0003-2604-0134}}
\author[HU]{M.E.~Christy\orcidlink{0000-0002-6350-9766}}
\author[JLAB]{S.~Covrig\orcidlink{0000-0001-9117-8493}}
\author[JLAB]{M.~M.~Dalton\orcidlink{0000-0001-9204-7559}}
\author[MAN]{W.~Deconinck\orcidlink{0000-0003-4033-6716}}
\author[JLAB]{M.~Diefenthaler\orcidlink{0000-0002-4717-4484}}
\author[TEMP]{B.~Duran\orcidlink{0000-0003-2344-5953}}
\author[MSU]{D. Dutta\orcidlink{0000-0002-7103-2849}}
\author[SUNO]{M. Elaasar\orcidlink{0009-0009-6688-1043}}
\author[JLAB]{R.~Ent\orcidlink{0000-0001-7015-2534}}
\author[JLAB]{H.~Fenker\orcidlink{0009-0004-9420-4705}}
\author[UCONN]{E.~Fuchey\orcidlink{0000-0003-0100-6052}}
\author[GLASGOW]{D.~Hamilton\orcidlink{0000-0001-5971-8947}}
\author[JLAB]{J.-O.~Hansen\orcidlink{0000-0002-7908-3886}}
\author[ODU]{F.~Hauenstein\orcidlink{0000-0002-1265-2212}}
\author[TEMP]{S.~Jia\orcidlink{0009-0001-8068-729X}}
\author[JLAB]{M.K.~Jones\orcidlink{0000-0002-7089-6311}}
\author[ANL]{S.~Joosten\orcidlink{0000-0003-4947-877X}}
\author[MSU]{M.L.~Kabir\orcidlink{0000-0003-2142-5166}}
\author[MSU]{A.~Karki\orcidlink{0000-0001-7650-2646}}
\author[JLAB]{C.~Keppel\orcidlink{0000-0002-7516-8292}}
\author[BOULDER]{E.~Kinney\orcidlink{0000-0002-4176-5283}}
\author[HU]{N.~Lashley-Colthirst\orcidlink{0000-0001-7695-1388}}
\author[WM,SBU]{W.B.~Li\orcidlink{0000-0002-8108-8045}}
\author[JLAB]{D.~Mack\orcidlink{0009-0004-3862-3307}}
\author[JLAB]{S.~Malace}
\author[JLAB]{M.~McCaughan\orcidlink{0000-0003-2649-3950}}
\author[ANL,TEMP]{Z.E.~Meziani\orcidlink{0000-0001-9450-2914}}
\author[JLAB]{R.~Michaels\orcidlink{0009-0008-1906-6313}}
\author[GLASGOW]{R.~Montgomery\orcidlink{0000-0002-2007-6833}}
\author[CUA]{M.~Muhoza}
\author[SACLAY]{C.~Mu\~noz~Camacho\orcidlink{0000-0002-0915-1740}}
\author[JMU]{G.~Niculescu\orcidlink{0000-0001-7640-4405}} 
\author[JMU]{I.~Niculescu\orcidlink{0000-0003-4526-7457}}
\author[REG]{Z.~Papandreou\orcidlink{0000-0002-5592-8135}}
\author[SBU]{S.~Park\orcidlink{0000-0002-8898-1231}}
\author[JLAB]{E.~Pooser\orcidlink{0000-0002-2107-3175}}
\author[TEMP]{M.~Rehfuss\orcidlink{0000-0001-5337-9550}}
\author[JLAB]{B.~Sawatzky\orcidlink{0000-0002-5637-0348}}
\author[JLAB]{G.R.~Smith\orcidlink{0000-0002-1755-809X}}
\author[JLAB]{H. Szumila-Vance\orcidlink{0000-0003-2548-4016}}
\author[REG]{A.~Teymurazyan\orcidlink{0000-0002-5970-5692}}
\author[YER]{H.~Voskanyan\orcidlink{0000-0001-9515-3568}}
\author[JLAB]{B.~Wojtsekhowski\orcidlink{0000-0002-2160-9814}}
\author[JLAB]{S.A.~Wood\orcidlink{0000-0002-1909-5287}}
\author[ANL]{Z.~Ye\orcidlink{0000-0002-1873-2344}}
\author[FIU]{C.~Yero\orcidlink{0000-0003-2822-7373}}
\author[UVA]{J.~Zhang\orcidlink{0000-0003-2610-904X}}
\author[UVA]{X.~Zheng\orcidlink{0000-0001-7300-2929}}

\affiliation[REG]{organization={University of Regina},
            city={Regina},
            postcode={S4S~0A2}, 
            state={SK},
            country={Canada}}
            
\affiliation[JLAB]{organization={Thomas Jefferson National Accelerator Facility},
            city={Newport News},
            postcode={23606}, 
            state={Virginia},
            country={USA}}
            
\affiliation[CUA]{organization={Catholic University of America},
            city={Washington},
            postcode={20064}, 
            state={DC},
            country={USA}}
            
\affiliation[YORK]{organization={University of York},
            city={Heslington},
            postcode={YO10~5DD}, 
            state={York},
            country={United Kingdom}}
            
\affiliation[FIU]{organization={Florida International University}, 
            city={University Park},
            postcode={33199}, 
            state={Florida},
            country={USA}}
            
\affiliation[OHIO]{organization={Ohio University},
            city={Athens},
            postcode={45701}, 
            state={Ohio},
            country={USA}}
            
\affiliation[KAU]{organization={Research Institute of Basic Sciences, Korea Aerospace University},
            city={Goyang},
            postcode={10540}, 
            country={Korea}}  
            
\affiliation[YONSEI]{organization={Department of Physics, Yonsei University},
            city={Wonju},
            postcode={26493}, 
            country={Korea}}   
            
\affiliation[ZAG]{organization={University of Zagreb},
            city={Zagreb},
            postcode={10000}, 
            country={Croatia}}
            
\affiliation[TEMP]{organization={Temple University},
            city={Philadelphia},
            postcode={19122}, 
            state={Pennsylvania},
            country={USA}}
            
\affiliation[ANL]{organization={Argonne National Laboratory},
            city={Lemont},
            postcode={60439}, 
            state={Illinois},
            country={USA}}
            
\affiliation[WM]{organization={College of William \& Mary},
            city={Williamsburg},
            postcode={23185}, 
            state={Virginia},
            country={USA}}
            
\affiliation[MSU]{organization={Mississippi State University},
            city={Mississippi State},
            postcode={39762}, 
            state={Mississippi State},
            country={USA}}
            
\affiliation[HU]{organization={Hampton University},
            city={Hampton},
            postcode={23669}, 
            state={Virginia},
            country={USA}}
               
\affiliation[CNU]{organization={Christopher Newport University},
            city={Newport News},
            postcode={23606}, 
            state={Virginia},
            country={USA}}

\affiliation[UVA]{organization={University of Virginia},
            city={Charlottesville},
            postcode={22903}, 
            state={Virginia},
            country={USA}}

\affiliation[MAN]{organization={University of Manitoba},
            city={Winnipeg},
            postcode={R3T~2N2}, 
            state={Manitoba},
            country={Canada}}
            
\affiliation[SUNO]{organization={Southern University at New Orleans},
            city={New Orleans},
            postcode={70126}, 
            state={Louisiana},
            country={USA}}

\affiliation[UCONN]{organization={University of Connecticut},
            city={Storrs},
            postcode={06269}, 
            state={Conneticut},
            country={USA}}
            
\affiliation[GLASGOW]{organization={University of Glasgow},
            city={Glasgow},
            postcode={G12~8QQ}, 
            country={United Kingdom}}
            
\affiliation[ODU]{organization={Old Dominion University},
            city={Norfolk},
            postcode={23529}, 
            state={Virginia},
            country={USA}}   
            
\affiliation[BOULDER]{organization={University of Colorado},
            city={Boulder},
            postcode={80309}, 
            state={Colorado},
            country={USA}}

\affiliation[SBU]{organization={Stony Brook University},
            addressline={}, 
            city={Stony Brook},
            postcode={11794}, 
            state={New York},
            country={USA}}

\affiliation[SACLAY]{organization={Universite Paris-Saclay, CNRS, IJCLab},
            city={Orsay},
            postcode={91406}, 
            country={France}}

\affiliation[JMU]{organization={James Madison University},
            city={Harrisonburg},
            postcode={22807}, 
            state={Virginia},
            country={USA}}

\affiliation[YER]{organization={A.I. Alikhanyan National Science Laboratory (Yerevan Physics Institute)},
            city={Yerevan},
            postcode={0036}, 
            country={Armenia}}
            
\begin{abstract}
Deep exclusive meson production (DEMP) reactions, such as $p(\vec{e},e'\pi^+)n$, provide opportunities to study the three-dimensional structure of the nucleon through differential cross section and beam- and target-spin asymmetry measurements. This work aims to probe the onset of the hard/soft factorization regime through the exclusive $p(\vec{e},e'\pi^+)n$ reaction, as measured in the KaonLT experiment at Jefferson Lab Hall C. A 10.6 GeV longitudinally polarized electron beam was incident on an unpolarized liquid hydrogen target, and the scattered electron and produced meson were detected in two magnetic focusing spectrometers, enabling precision cross section measurements. The cross section ratio $\sigma_{LT'}/\sigma_0$ was extracted from the beam-spin asymmetry $A_{LU}$. The $t$-dependence of $\sigma_{LT'}/\sigma_0$ was determined at fixed $Q^2$ and $x_B$ over a range of kinematics from $2<Q^2<6$ GeV$^2$ above the resonance region ($W>2$ GeV). Furthermore, these data are combined with recent results from CLAS/CLAS12 to determine the $Q^2$-dependence of $\sigma_{LT'}/\sigma_0$ at two ($x_B$, $t$) settings. This was fairly flat, with $Q^2$ not having a measurable effect on the value of  $\sigma_{LT'}/\sigma_0$ in the range explored. Results are compared to predictions from the generalized parton distribution (GPD) formalism, which relies explicitly on hard/soft factorization, and Regge formalism. The Regge models better predict $\sigma_{LT'}/\sigma_0$, which suggests that the factorization regime is not yet reached.
\end{abstract}


\end{frontmatter}

\section{Introduction}
\label{introduction}

A quantitative description of simple hadronic systems such as light mesons and nucleons is essential to our understanding of nuclear matter. Deep exclusive meson production (DEMP) reactions, such as $p(\vec{e},e'\pi^+)n$, provide opportunities to study the three-dimensional structure of the nucleon through differential cross section and beam- and target-spin asymmetry measurements. DEMP reactions can be conveniently described using three Lorentz invariants. $Q^2=-(p_e-p_{e'})^2$ is the negative of the four-momentum transfer squared of the virtual photon. Additionally, the reaction is characterized by the invariant mass of the virtual photon-nucleon system, $W^2=(p_p+p_{\gamma^*})^2$, and the Mandelstam variable $t=(p_{\gamma^*}-p_{\pi})^2$. Alternatively, the Bjorken scaling variable $x_B=Q^2/(2p_p\cdot p_{\gamma^*})$ may replace $W$.

Hard/soft factorization describes the expression of the $\gamma^*p$ amplitude as the convolution of a hard-scattering subprocess and a non-perturbative (soft) subprocess. A factorization theorem has been proven for DEMP events involving longitudinally polarized virtual photons \citep{Radyushkin_1996, Collins_1997}, and the contribution of transversely polarized virtual photons has been treated as a twist-3 effect in this approach \citep{Goloskokov_2009}. Hard/soft factorization is expected to apply in the limit of large $Q^2$ at fixed $x_B$ and $t$. For Deep Virtual Compton Scattering (DVCS), factorization appears valid even at modest $Q^2\approx 2$ GeV$^2$ \cite{Camacho_2006, Girod_2008}, but the  minimum $Q^2$ for which factorization may be valid for DEMP is still unknown \cite{Favart_2016}.

The identification of this factorization regime for $p(\vec{e},e'\pi^+)n$ is of high interest to hadronic physics, as factorization allows for the extraction of Generalized Parton Distributions (GPDs). GPDs \citep{Muller_2014, Diehl_2003} unify the concepts of parton distributions and hadronic form factors by correlating the transverse position and longitudinal momentum of partons. Measuring these observables is expected to facilitate numerous advances in our understanding of nucleon structure, for example providing information on the orbital angular momentum of partons, which is needed for solving the proton spin crisis \cite{Diehl_2003,Ji_1997}. DEMP reactions provide complementary information to DVCS towards the extraction of GPDs; DVCS primarily probes chiral-even GPDs, whereas DEMP also probes chiral-odd GPDs \cite{Goloskokov_2009,Goldstein_2015}. The DEMP reaction $p(\vec{e},e'\pi^+)n$ in particular has a significant contribution from the GPD $H_T$ \citep{Goloskokov_2009}, therefore polarized $\pi^+$ observables in the factorization regime could be used to probe fundamental quantities such as the still unknown tensor charge of the nucleon, which is calculated from the integral of $H_T$ \citep{Goldstein_2015}.

To investigate the onset of hard/soft factorization, the KaonLT experiment (E12-09-011 \cite{Horn_2008}) at Hall C of the Thomas Jefferson National Accelerator Facility (Jefferson Lab or JLab) measured DEMP reactions over a range of kinematics from $2<Q^2<6$ GeV$^2$ above the resonance region ($W>2$ GeV). The KaonLT data will allow for the extraction of a number of hadronic structure observables including the total cross section $\sigma_0$, longitudinal and transverse cross sections $\sigma_L$ and $\sigma_T$ (where the subscript denotes the virtual photon polarization), and interference cross sections $\sigma_{LT}$, $\sigma_{TT}$, and $\sigma_{LT'}$.

\begin{figure}[h]
    \centering
    \includegraphics[width=0.8\linewidth]{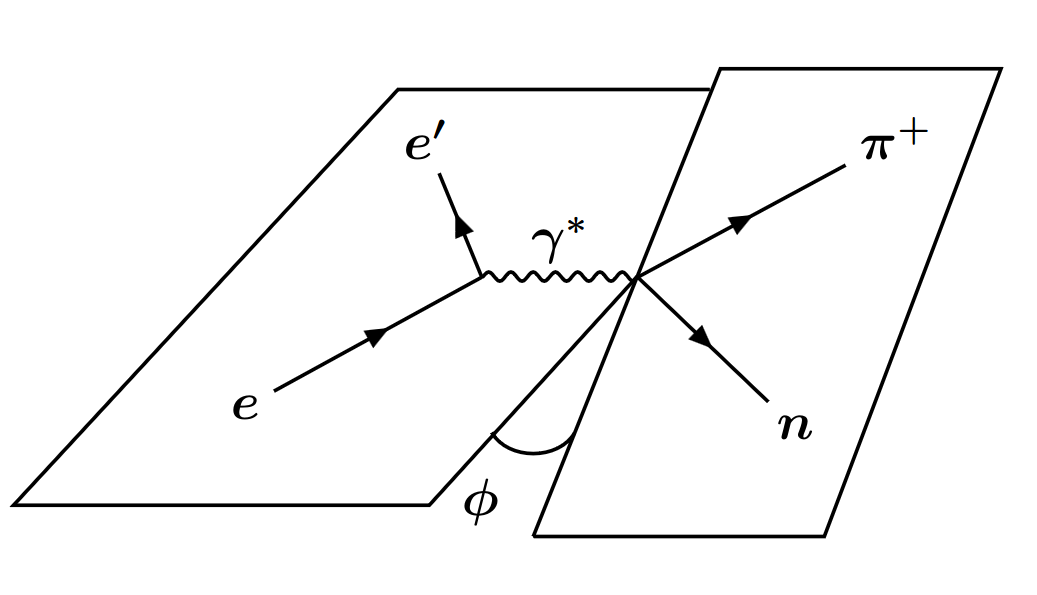}
    \caption{Reaction diagram for $p(\vec{e},e'\pi^+)n$. The angle $\phi$ is defined as the azimuthal angle between the electron scattering plane (defined by $e$ and $e'$) and the hadron reaction plane (defined by $\pi^+$ and $n$).}
    \label{fig:phi}
\end{figure}

In this work, the cross section ratio $\sigma_{LT'}/\sigma_0$ is extracted from beam-spin asymmetry measurements of $p(\vec{e},e'\pi^+)n$. The prime in the subscript of $\sigma_{LT'}$ denotes polarization, as $\sigma_{LT'}$ is only accessible in the case of a longitudinally polarized incident electron beam. $\sigma_{LT'}$ is proportional to the imaginary part of interference between longitudinally and transversely polarized virtual photons (as opposed to $\sigma_{LT}$, which is accessible with an unpolarized beam, and is proportional to the real part of the same interference amplitude) \cite{Diehl_2005}. In the one-photon exchange approximation, this asymmetry can be expressed as \citep{Diehl_2005,Arens_1997}
\begin{multline}
    A_{LU}(Q^2,x_B,t,\phi) = \\
    \frac
    {\sqrt{2\epsilon(1-\epsilon)} \frac{\sigma_{LT'}}{\sigma_0}\sin\phi}
    {1+\sqrt{2\epsilon(1+\epsilon)}\frac{\sigma_{LT}}{\sigma_0}\cos\phi
    +\epsilon\frac{\sigma_{TT}}{\sigma_0}\cos2\phi},
    \label{eqn:functional}
\end{multline}
where $\epsilon$ is the ratio of longitudinal to transverse polarized virtual photon flux and $\phi$ is the azimuthal angle shown in Fig.~\ref{fig:phi} \cite{Bacchetta_2004}. All three interference terms are required to vanish when $t=-|t|_{\textrm{min}}$ and $t=-|t|_{\textrm{max}}$; for these values the $\gamma^*p\rightarrow \pi^+n$ reaction is collinear in the struck proton rest system and $\phi$ is undefined. The subscript $LU$ specifies the asymmetry resulting from a longitudinally polarized incident electron beam and an unpolarized target. $\sigma_{LT'}/\sigma_0$ is extracted from the asymmetry via the $\sin \phi$ amplitude of $A_{LU}$, defined as $A_{LU}^{\sin\phi}=\sqrt{2\epsilon(1-\epsilon)}\sigma_{LT'}/\sigma_0$.

\begin{figure}
\includegraphics[width=1.0\linewidth]{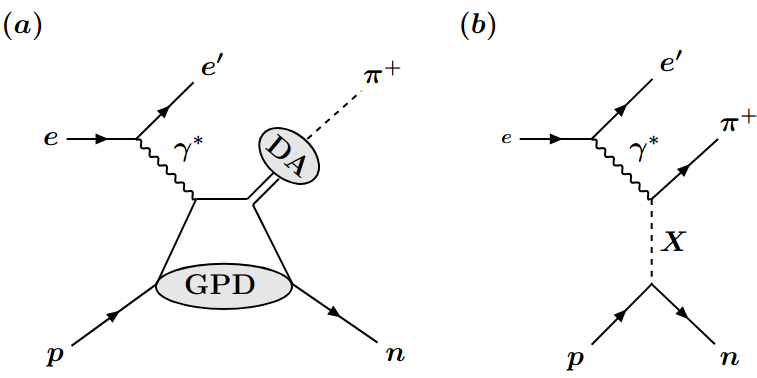}
    \caption{\label{fig:models} Exclusive $\pi^+$ electroproduction from the proton. (a) Factorization of the reaction into a hard scattering part and a soft part described by a GPD. An additional soft part known as the pion distribution amplitude (DA) describes the final state pion formation. (b) A Regge process, in which $X$ represents the exchange of several particles along a Regge trajectory up to a cutoff. }
\end{figure}

There are two main approaches to describe this observable: the first is to assume factorization and describe the reaction using GPDs (Fig.~\ref{fig:models}(a)). An alternative description of DEMP reactions is based on Regge models. Here, the interaction is mediated by the exchange of meson trajectories in the $t$ channel (Fig.~\ref{fig:models}(b)). Regge models \citep{Regge_1959} have been extended from photoproduction ($Q^2$=0) \citep{Yu_2011} to DEMP \citep{Laget_2020}, and have successfully described DEMP reactions at Hall C kinematics \citep{Huber_2008}. Unlike GPDs, the validity of Regge models does not explicitly rely on hard/soft factorization, neither do they describe the three-dimensional structure of the nucleon.
This work compares $\sigma_{LT'}/\sigma_0$ to predictions from three models \cite{Yu_2011, Goloskokov_2011, Vrancx_2014} to explore if a GPD or Regge description is more applicable to DEMP reactions at these kinematics.  If the GPD-based model clearly outperforms the Regge models, it would be a strong indication of factorization validity at these kinematics.
 
\section{Experiment}

$A_{LU}$ is experimentally calculated as a fractional difference of yield based on the helicity of the incident electron $Y^\pm$.
\begin{equation}
    A_{LU} = \frac{1}{P}\bigg(\frac{Y^+-Y^-}{Y^++Y^-}\bigg)
    \label{eqn:bsa}
\end{equation}
$A_{LU}$ has been previously measured above the resonance region at Jefferson Lab Hall B
in exclusive $\pi^+$ production \citep{Diehl_2023,Diehl_2020}, and in exclusive $\pi^0$ \citep{DeMasi_2008}. This work reports the first measurement of $A_{LU}$ in $p(\vec{e},e'\pi^+)n$ from Hall C as part of the KaonLT experiment.
The experimental setup of Hall C, with two small acceptance and high precision spectrometers, enables a measurement
with significantly finer kinematic binning and cleaner identification of the exclusive final state compared to previous measurements of this observable.

The KaonLT experiment ran for 90 days, of which 40 were used for the high beam energy data set reported here.
A continuous wave electron beam with energy 10.585 GeV and current up to 70 $\mu$A was used. The beam energy was determined to $\pm$3.6 MeV by measuring the bend angle of the beam into Hall C, as it traversed a set of dipole magnets with precisely calibrated field integrals \citep{Higinbotham_2013}. The beam helicity was flipped at a frequency of 30 Hz in a pseudo-random sequence, with a helicity-correlated charge asymmetry of up to 0.1\% \citep{Benesch_2018}. No dedicated beam polarization measurements were made in Hall C. Rather, Mott polarimetry measurements were taken at the injector to the accelerator ($90\pm1\%$) \citep{Grames_2020}, and a calculation of the spin precession through the accelerator indicated that for this beam energy Hall C receives 99\% of the source polarization. These gave a result of $89^{+1}_{-3}\%$ longitudinal beam polarization to Hall C, where the uncertainty is determined from the beam energy uncertainty and the range of possible linac energy imbalance. This spin-precession calculation is consistent with direct measurements in Hall B, which gave on average $86\pm2\%$ polarization during the same run period.
Beam quality was assured by continuous measurements from three beam position monitors \cite{Benesch_2022}, four beam current monitors \cite{Mack_1992}, and an Unser monitor \cite{Unser_1992}. 

\begin{figure}
    \centering
    \includegraphics[width=1.0\linewidth]{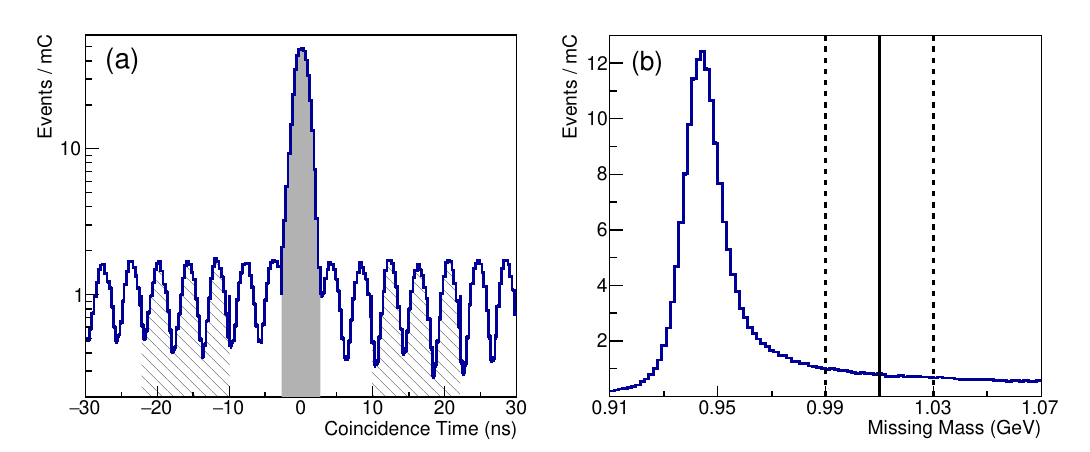}
    \caption{Coincidence time and missing mass spectra for Q$^2$=3.0 GeV$^2$, $x_B$=0.25, center SHMS setting. (a) Coincidence time between the HMS and SHMS. The prompt peak selected is highlighted in grey, and the windows used to subtract random coincidences are hatched. (b) Missing mass distribution of $p(\vec{e},e'\pi^+)n$. The solid line shows the upper missing mass cut used, and the dashed lines show the variation of the cut used to calculate a cut dependence. The lower missing mass cut is 0.91 GeV, and its contribution to the cut dependence is evaluated by removing this cut entirely.}
    \label{fig:mmcoin}
\end{figure}

The electron beam was incident upon a 10 cm (762 mg/cm$^2$) cryogenic unpolarized liquid hydrogen target. Charged pions were detected in the recently commissioned Super High Momentum Spectrometer (SHMS), which has momentum acceptance $\Delta p/p$ from -10 to +20\% of the central momentum, a maximum central momentum of 11 GeV/c, and covers a solid angle of $\Delta\Omega=4$ msr \citep{Ali_2024}. Scattered electrons were detected in the High Momentum Spectrometer (HMS), which has momentum acceptance $\Delta p/p=\pm 8\%$ and solid angle acceptance $\Delta\Omega=7$ msr. The maximum HMS central momentum is 7 GeV/c, and the central momentum is chosen to set the values ($Q^2$,$x_B$) for each experimental setting \citep{Blok_2008}. Both spectrometers include two drift chambers for track reconstruction, hodoscope arrays for triggering, threshold Cherenkov detectors and lead-glass calorimeters for particle identification.  An open trigger in the hadron arm resulted in the collection of high-quality $\pi^+$ data in addition to the $K^+$ sample.
Positive pions were identified in the SHMS using an aerogel Cherenkov detector with refractive index $n=1.015$ (for $p_{\pi}<5$ GeV/c) or $n=1.011$ (for $p_{\pi}>5$ GeV/c), for a pion detection efficiency of 97\%. Electrons were identified in the HMS via a gas Cherenkov detector filled with C$_4$F$_{10}$ at 0.48 atm (refractive index 1.0008) in combination with the lead-glass calorimeter.

A coincidence window of $\pm$2.25 ns between the SHMS and HMS trigger was used to ensure both the pion and electron came from the same beam bunch.
Any remaining contamination from real $e-p$ and $e-K^+$ coincidences was eliminated with a missing mass cut of $0.91<m_x<$1.01 GeV (Fig.~\ref{fig:mmcoin}). For the $p(\vec{e},e'\pi^+)n$ reaction, the reconstructed missing mass, $m_x^2=(m_p+p_e-p_{e'}-p_{\pi})^2$, is close to the free neutron mass, with a radiative tail extending to higher $m_x$ (Fig.~\ref{fig:mmcoin}). The upper cut on the missing mass was selected to include as much of the radiative tail as possible without including contamination from $e-K^+$ events or Semi-Inclusive Deep Inelastic Scattering (SIDIS), which begins at $m_x=1.05$ GeV (considering the experimental resolution of $m_x$). To investigate the effects of radiation, events were simulated both with and without radiative corrections applied in the Hall C Monte Carlo SIMC \cite{Ent_2001}. The simulated asymmetry was calculated for every experimental ($Q^2$,$x_B$,$-t$) bin, and it was found that the radiative corrections changed the input asymmetry by no more than 0.6\%. As this is significantly smaller than the other uncertainties, no radiative corrections were applied to the data.

A sample of random coincidences ($\sim$3\% of events at $x_B$=0.4 and $\sim$12\% at $x_B$=0.25) was subtracted before the data was beam charge-normalized. Additionally, a comparable amount of data was collected on two aluminum foils placed 10 cm apart. This data was also beam charge-normalized, then subtracted from the experimental yields to remove background from the aluminum target cell walls (1--2\% of events). The detector efficiencies were verified to be uncorrelated with the electron beam helicity, and therefore they cancel in the calculation of $A_{LU}$ (Eqn. \ref{eqn:bsa}).

\begin{figure}[h!]
\includegraphics[width=1.0\linewidth]{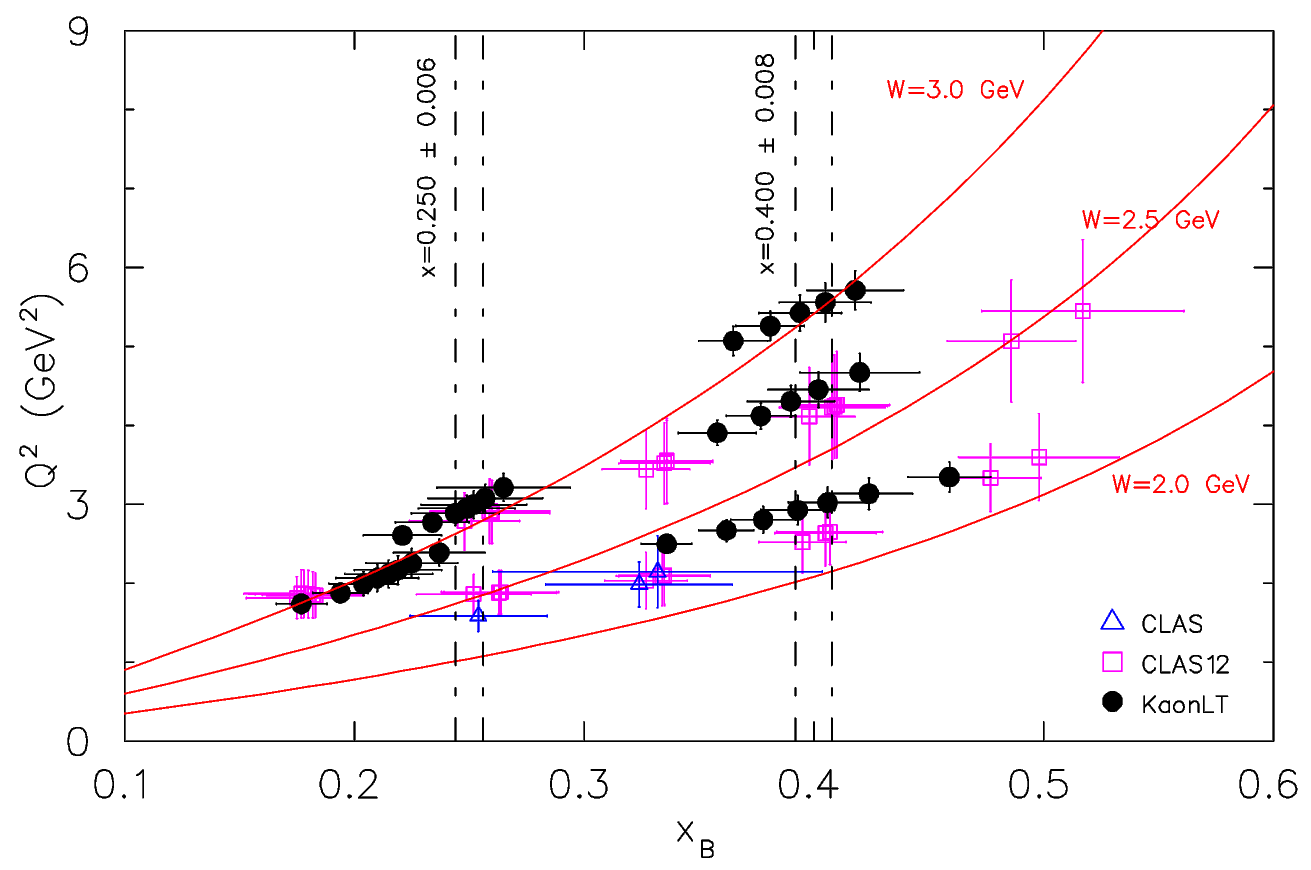}
\caption{\label{fig:phasespace} (Color online) Phase space plot of the kinematics for which $\sigma_{LT'}/\sigma_0$ has been measured \citep{Diehl_2023, Diehl_2020} [This work]. The error bars indicate the standard deviations in $x_B$ and $Q^2$ spanned by each data point. Note that 3 variables $(Q^2,x_B,t)$ are required to fully describe exclusive reaction kinematics, the different data points are taken at  different $t$-values not indicated here. Each grouping of data points represents one setting, where $-t$ increases from left to right. For both CLAS datasets, only $-t<0.8$ GeV$^2$ data are shown, corresponding to the upper range of the KaonLT data. By combining these data sets, the $Q^2$ dependence of $\sigma_{LT'}/\sigma_0$ can be determined at fixed $-t$ for two values of $x_B$, shown as dashed lines.}

\end{figure}

\begin{figure*}[hbt!]
\centering 
\includegraphics[width=0.9\linewidth]{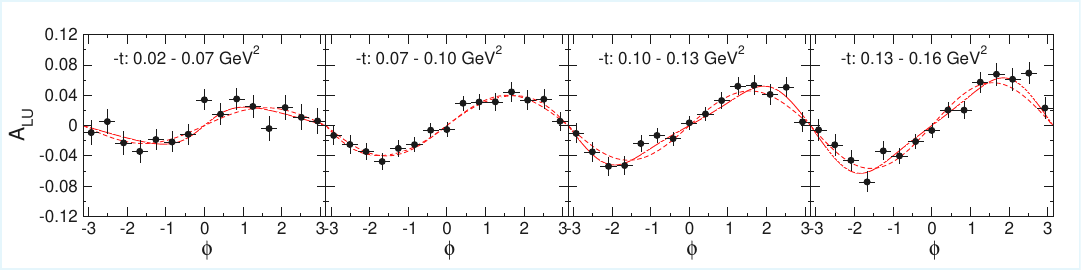}
\caption{\label{fig:asymmetry} $A_{LU}$ as a function of $\phi$ for the first four $t$-bins for central values of $Q^2=3$ GeV$^2$, $x_B=$0.25. The solid line shows the data fit with Eqn.~\ref{eqn:functional} and the dashed line Eqn.~\ref{eqn:approx} (see text for explanation). Uncertainties are statistical only.}
\end{figure*}

\begin{figure*}[hbt!]
\centering
\includegraphics[width=1.0\linewidth]{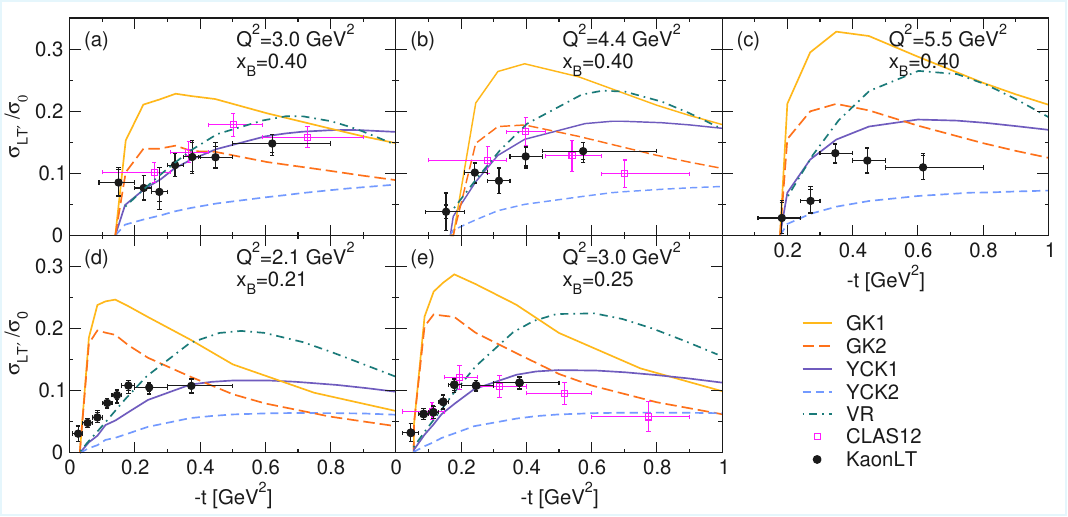}
\caption{\label{fig:results} (Color online) The extracted $\sigma_{LT'}/\sigma_0$ as a function of $-t$ for each $(Q^2,x_B)$ setting. The horizontal error bar indicates the width of the $t$-bin, and the double vertical error bar shows the statistical and total errors. Exact values of $\sigma_{LT'}/\sigma_0$ and its uncertainties, as well as binned kinematics, are presented in Table \ref{tab:test}. Settings are labeled by the central $(Q^2,x_B)$ values, but the mean values $<Q^2>$ and $<x_B>$ vary per bin. The smooth curves represent theory predictions (see legend). GK1 refers to the default GK version \cite{Berthou_2015}, and GK2 is the GK model with the modification of $H_T \rightarrow H_T*2$, following the example of Ref. \citep{Diehl_2023}. YCK1 is the YCK model with the nucleon EMFFs parametrized with GPDs, whereas YCK2 uses a dipole parametrization. The VR model is also shown, and all models are described in the text. At low $-t$, where $\sigma_{LT'}/\sigma_0$ varies rapidly, the models are evaluated at the mean kinematics per bin of the KaonLT data. When the plotting range extends beyond the $-t$ range of the data, the model is evaluated using the kinematics of the highest-$t$ KaonLT point available. In addition, some CLAS12 data at similar kinematics are plotted for comparison. CLAS12 data were chosen based on two criteria; first, the mean kinematics must be similar, and second, there must be significant overlap between the limits of the kinematic bins (see Fig.~\ref{fig:phasespace}.).}
\end{figure*}

\begin{figure}[htb]
\includegraphics[width=1.0\linewidth]{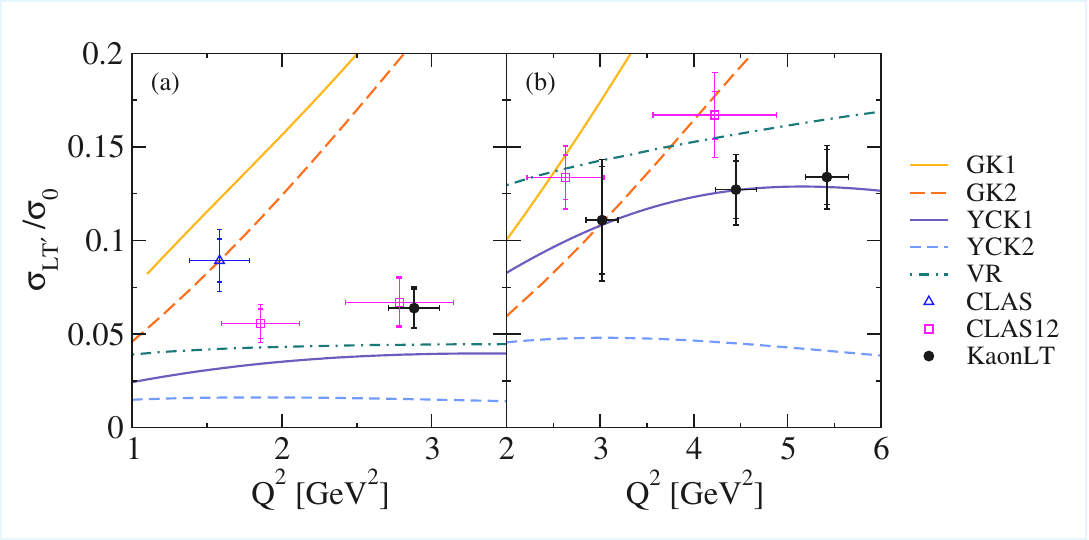}
\caption{\label{fig:q2} (Color online) Values of $\sigma_{LT'}/\sigma_0$ from three experiments \citep{Diehl_2023,Diehl_2020} [This work] plotted as a function of $Q^2$ for the kinematics (a) $x_B=0.250\pm 0.006$, $-t=0.111\pm0.004$ GeV$^2$, (b) $x_B=0.400\pm0.008$, $-t=0.37\pm0.03$ GeV$^2$. The $Q^2$-dependence of the data is consistent, within error, with a horizontal line. The GK1 and GK2 curves continue a uniform increase after going off-scale. Theory curves are evaluated at (a) $\epsilon=0.71$ (b) $\epsilon=0.77$. Fig.~\ref{fig:phasespace} shows the $x_B$ regions selected for this analysis. Only the points which also satisfy the criterion in $-t$ are included here.}
\end{figure}

The ($Q^2,x_B$) settings studied in this experiment are shown in Fig.~\ref{fig:phasespace}. 
For each ($Q^2,x_B$) setting, the data were split into 5---8 bins in $t$ and 15 bins in $\phi$, with the number of $t$-bins determined by the number of events at each setting. 
In exclusive pion production, the experimental acceptances in $x_B$, $Q^2$ and $t$ are correlated. Thus, for each $t$-bin (but independent of $\phi$), the mean $\langle Q^2 \rangle$ and $\langle x_B \rangle$ values of the data vary from the `central' values. This is visible in Fig.~\ref{fig:phasespace}, where each setting is not a single point, but a group of points, each representing one $t$-bin. 
The value of $\epsilon$, which determines the conversion from $A_{LU}^{\sin\phi}$ to $\sigma_{LT'}/\sigma_0$, varies by at most 0.015 within a setting. See Table \ref{tab:test} for the precise kinematics of each measurement.

For each ($Q^2,x_B$) setting, the HMS angle and central momentum, as well as the SHMS central momentum, were kept fixed. To attain full coverage in $\phi$, data were taken with the SHMS at $\pm 3^{\circ}$ of the $\vec{q}$-vector direction (virtual photon momentum), in addition to data centered on the $\vec{q}$-vector (center setting). 
The asymmetry was calculated according to Eqn.~\ref{eqn:bsa} for each $t$-bin at each of the three SHMS angles. An error-weighted average was then taken to obtain a complete $\phi$ distribution.
Fig.~\ref{fig:asymmetry} shows the binned asymmetry for central kinematics of $Q^2$=3 GeV$^2$, $x_B$=0.25 for illustration.

The relevant kinematic variables, $Q^2$, $x_B$, $W$, and $t$ were reconstructed from the measured spectrometer quantities. Using proton-electron elastic coincidences (where the proton is detected in the SHMS and the electron in the HMS) as an over-determined reaction, the beam momentum and the spectrometer central momenta were determined absolutely to $<0.5\%$, while the incident beam angle and spectrometer central angles were absolutely determined to $<0.5$ mrad (method in \cite{Volmer_2000}).

Previous work assumed that $\sigma_{TT}/\sigma_0 \ll 1$ and $\sigma_{LT}/\sigma_0 \ll 1$, such that Eqn.~\ref{eqn:functional} simplifies to
\begin{equation}
    A_{LU} = \sqrt{2\epsilon(1-\epsilon)}\frac{\sigma_{LT'}}{\sigma_0}\sin\phi,
    \label{eqn:approx}
\end{equation}
the justification being that Eqns.~\ref{eqn:functional} and \ref{eqn:approx} gave extremely similar results for $A_{LU}^{\sin \phi}=\sqrt{2\epsilon(1-\epsilon)}\sigma_{LT'}/\sigma_0$ \cite{Diehl_2023, Diehl_2020}. Neglecting $\sigma_{LT}/\sigma_0$ and $\sigma_{TT}/\sigma_0$ appears to be a low $-t$ approximation, which is insufficiently accurate for our fits at higher $-t$, as seen in the last panel of Fig.~\ref{fig:asymmetry}. The authors are aware of no theoretical constraints for why $\sigma_{LT}/\sigma_0$ and $\sigma_{TT}/\sigma_0$ should be negligible. Additionally, a Monte Carlo study was performed which determined that at the experimental precision, it is not feasible to accurately determine if $\sigma_{LT}/\sigma_0$ and $\sigma_{TT}/\sigma_0$ are negligible.
Therefore, $A_{LU}^{\sin\phi}$ was determined using Eqn.~\ref{eqn:functional}, with $\sigma_{LT}/\sigma_0$ and $\sigma_{TT}/\sigma_0$ left as free parameters in the fit. The statistical error on $A_{LU}^{\sin\phi}$ is taken as the error of fitting when including the statistical uncertainties per $\phi$ bin. The cross section ratio $\sigma_{LT'}/\sigma_0$ and its statistical uncertainty were then extracted from $A_{LU}^{\sin\phi}$.

There are three main sources of systematic uncertainty. First is the difference in $\sigma_{LT'}/\sigma_0$ obtained using Eqn.~\ref{eqn:functional} and Eqn.~\ref{eqn:approx}. Since such a difference is unidirectional, the total systematic error (obtained from the quadrature sum of systematic uncertainties) is asymmetric, denoted $\delta_{\textrm{sys}}^\uparrow$ and $\delta_{\textrm{sys}}^\downarrow$ for the upper and lower error bars.
This is the dominant systematic, contributing an average error of 12\%, but up to 70\% in the bin $(Q^2, x_B)$=(3.0, 0.40) and $0.25<-t<0.3$. Additionally, the uncertainty on the beam polarization contributed an uncertainty of ${}^{+1.1}_{-3.4}\%$, and the dependence of $\sigma_{LT'}/\sigma_0$ on the exact values used in the coincidence time and missing mass cuts contributed between 1--7\%, with one outlier at 12\%. In general, the systematics are on par with or smaller than the statistical error, and of a similar magnitude to the systematic errors in the CLAS12 measurement \cite{Diehl_2023}.

\section{$t$-dependence comparison with theoretical expectations}

Fig.~\ref{fig:results} shows $\sigma_{LT'}/\sigma_0$ versus $-t$ in comparison to theoretical predictions and CLAS12 data. The results from this work are in good agreement with results from CLAS12, showing a similar magnitude and $t$-dependence of $\sigma_{LT'}/\sigma_0$ \citep{Diehl_2023}.  In particular, KaonLT and CLAS data points with very similar $Q^2$, $x_B$ and $t$ agree within the quoted uncertainties.

Three distinct models were considered, of which two are based on Regge trajectories, and one on GPD formalism. The Goloskokov-Kroll (GK) model \citep{Goloskokov_2009, Goloskokov_2011} calculates $\sigma_{LT'}$ for deep exclusive $\pi^+$ production in terms of the twist-2 longitudinal ($\widetilde{E}, \widetilde{H}$) and twist-3 transverse ($E_T, H_T$) GPDs, with pion pole contributions at low $-t$. The default version of the GK model, denoted GK1, shows better agreement for $x_B=0.40$ than for $x_B=0.25$, in which case its $t$-dependence does not match the data. In Ref. \citep{Diehl_2023}, the argument was made that increasing the GPD $H_T$ in the GK calculation resulted in better agreement with experimental data. In this work, the curve GK2 is the GK model with the modification $H_T\rightarrow 2 H_T$. GK2 has a lower magnitude than GK1, bringing it closer to data, but it still does not re-create the $t$-dependence of $\sigma_{LT'}/\sigma_0$ at all kinematics.

Secondly, the Vrancx-Ryckebush (VR) model \citep{Vrancx_2014} considers Reggeized $\pi(140)$, $\rho(770)$, and $a_1(1260)$ exchanges. Including only $\pi(140)$ and $\rho(770)$ leads to a vanishing $A_{LU}$. The inclusion of the axial-vector $a_1(1260)$ exchange generates a non-zero $A_{LU}$ through interference with vector $\rho(770)$ exchange \citep{Kaskulov_2010}. However, this interference is still insufficient to reproduce $A_{LU}$ from previous CLAS data \citep{Avakian_2004} without proper treatment of the ``resonant effect'' caused by nucleon form factors. In this work, the VR model predicts the data reasonably well at low $-t$, but it does not capture the plateau of $\sigma_{LT'}/\sigma_0$ that occurs at higher $-t$.

Finally, the Yu-Choi-Kong (YCK) model predicts $A_{LU}$ using Regge propagators, with contribution of the magnetic moment term of the nucleon with the Pauli form factor $F_2(Q^2)$. It incorporates the exchange of tensor meson $a_2(1320)$ with axial mesons $a_1$ and $b_1(1235)$, which were not included in the earlier version \citep{Choi_2015}. In the new model, the electromagnetic form factors (EMFFs) of the nucleon are considered in two categories: the GPD-mediated form \citep{Guidal_2005}, designated YCK1, and the typical dipole form, designated YCK2. The YCK1 model presents the best agreement with this work. YCK2 underestimates $\sigma_{LT'}/\sigma_0$, but YCK1 provides a reasonable prediction of both the magnitude and $t$-dependence of $\sigma_{LT'}/\sigma_0$. 

Quantitatively, YCK1 has the lowest average $\chi^2/$NDF, although none of the calculated $\chi^2/$NDF values are close to 1. The VR model has the second-lowest $\chi^2/$NDF, but this is artificially influenced by the majority of data points being at low $-t$, before the VR model clearly diverges from the data.

\begin{table*}[htb!]
    \centering
    \begin{tabular}{c c c c c c c c c}
    \hline 
    \hline 
        Data set & $\langle -t \rangle$ [GeV${}^2$] & $\langle Q^2 \rangle$ [GeV${}^2$] & $\langle x_B \rangle$ &  $\langle \epsilon \rangle$ & $\sigma_{LT'}/\sigma_0$ & $\delta_{\textrm{stat}}$ & $\delta_{\textrm{sys}}^{\downarrow}$ & $\delta_{\textrm{sys}}^{\uparrow}$ \\
        \hline 
        \multicolumn{9}{c}{\rule{0pt}{3ex} (a) $x_B = 0.250 \pm 0.006$, $-t = 0.111 \pm 0.004$ GeV${}^2$} \\
        \rule{0pt}{3ex} CLAS & 0.108 & 1.58 $\pm$ 0.20 & 0.254 $\pm$ 0.030 & 0.648 & 0.0893 & 0.0115 & 0.0118 & 0.0118 \\
        CLAS12 & 0.111 & 1.86 $\pm$ 0.26 & 0.252 $\pm$ 0.025 & 0.866 & 0.0556 & 0.0078 & 0.0064 & 0.0064 \\
        CLAS12 & 0.111 & 2.78 $\pm$ 0.36 & 0.248 $\pm$ 0.024 & 0.661 & 0.0670 & 0.0126 & 0.0048 & 0.0048 \\
        KaonLT & 0.115 & 2.88 $\pm$ 0.17 & 0.244 $\pm$ 0.014 & 0.678 & 0.0653 & 0.0100 & 0.0034 & 0.0047 \\
        
        \multicolumn{9}{c}{\rule{0pt}{3ex} (b) $x_B = 0.400 \pm 0.008$, $-t = 0.37 \pm 0.03$ GeV${}^2$} \\
        \rule{0pt}{3ex} CLAS12 & 0.365 & 2.63 $\pm$ 0.41 & 0.405 $\pm$ 0.022 & 0.895 & 0.1337 & 0.0119 & 0.0119 & 0.0119 \\
        KaonLT & 0.376 & 3.02 $\pm$ 0.17 & 0.406 $\pm$ 0.022 & 0.885 & 0.1263 & 0.0233 & 0.0115 & 0.0103 \\
        CLAS12 & 0.398 & 4.22 $\pm$ 0.66 & 0.408 $\pm$ 0.023 & 0.704 & 0.1670 & 0.0187 & 0.0127 & 0.0127 \\
        KaonLT & 0.395 & 4.45 $\pm$ 0.22 & 0.402 $\pm$ 0.022 & 0.719 & 0.1272 & 0.0153 & 0.0108 & 0.0029 \\
        KaonLT & 0.347 & 5.42 $\pm$ 0.23 & 0.394 $\pm$ 0.018 & 0.525 &  0.1324 & 0.0143 & 0.0082 & 0.0030 \\
        \hline 
        \hline 
    \end{tabular}
    \caption{Data used in two scans of the Q$^2$-dependence of $\sigma_{LT'}/\sigma_0$ (Fig. \ref{fig:q2}) \cite{Diehl_2023}\citep{Diehl_2020}[This work]. Quantities enclosed in angled brackets denote the mean value of that variable in each $t$-bin, and the standard deviation of $Q^2$ and $x_B$ in the bin is provided. The columns $\delta_{\textrm{stat}}$, $\delta_{\textrm{sys}}^\downarrow$ and $\delta_{\textrm{sys}}^\uparrow$ refer to the uncertainties of $\sigma_{LT'}/\sigma_0$. In the selection of data points for each scan, $\langle x_B \rangle$ and $\langle -t \rangle$ were constrained to within the provided range, even if the standard deviation of the variable is larger.
    The systematic error of the KaonLT data is asymmetric, with $\delta_{\textrm{sys}}^\downarrow$ ($\delta_{\textrm{sys}}^\uparrow$) denoting the lower (upper) bounds. The total error in each direction is the sum of $\delta_{\textrm{stat}}$ and $\delta_{\textrm{sys}}^\downarrow$ ($\delta_{\textrm{sys}}^\uparrow$) in quadrature.}
    \label{tab:q2}
\end{table*}

\section{$Q^2$-dependence comparison with theoretical expectations}

By comparing data between CLAS, CLAS12, and this work, two kinematic ranges were identified where it was possible to hold $x_B$ and $t$ essentially constant while varying $Q^2$, allowing the $Q^2$-dependence of $\sigma_{LT'}/\sigma_0$ to be determined (Fig. \ref{fig:phasespace}). The results are shown in Fig.~\ref{fig:q2} and tabulated in Table \ref{tab:q2}.  Predictions of all models shown on Fig.~\ref{fig:results} are also computed for the kinematics of  Fig.~\ref{fig:q2}. To generate a smooth curve, a single $\epsilon$ value is used when calculating theoretical predictions; some deviations between data and theory could be a result of this averaging effect.
In Fig.~\ref{fig:results}, it can be seen that most theory curves incorporate a $Q^2$-dependence, in which the magnitude of the predicted $\sigma_{LT'}/\sigma_0$ increases with $Q^2$. This work suggests that in this regime, a description involving a significant $Q^2$-dependence is not supported by the data. 

At the two ($x_B,t)$ points investigated, $(x_B=0.400 \pm 0.008, -t=0.37 \pm 0.03)$ and $(x_B=0.250\pm0.006, -t=0.111\pm0.004)$, the asymmetry is largely independent of $Q^2$.
Future measurements of $\sigma_{LT'}/\sigma_0$ over a wider range of $Q^2$ would be beneficial to further test this dependence. Additional data have been taken in Hall C (PionLT experiment, E12-19-006 \cite{Gaskell_2009}) which could be used for this analysis. 

Both Regge-based models (VR, YCK) outperform the GPD-based GK model in predicting the $Q^2$-dependence of $\sigma_{LT'}/\sigma_0$.
The GK model is among the best available GPD models, and provides a good benchmark for comparison of experimental data with factorization expectations.
As such, the fact that the GK model provides a poor description of the $t$-dependence of $\sigma_{LT'}/\sigma_0$ at low $-t$ in Fig. \ref{fig:results}, as well as a poor description of the $Q^2$-dependence in Fig. \ref{fig:q2}, lends support to the hypothesis that the factorization regime has not yet been reached. However, it should be noted that $\sigma_{LT'}/\sigma_0$ is a twist-3 quantity, which adds additional complexity to a GPD-based prediction. Some of the deviations of the GK model from the data could owe to higher-twist effects not included in the GK model while still being consistent with factorization. We encourage the development of future GPD models which could be used for a more thorough test of $\sigma_{LT'}/\sigma_0$. 

\section{Discussion and conclusions}

The recent CLAS12 measurement of $\sigma_{LT’}/\sigma_0$ \cite{Diehl_2023} concluded that the GK2 model best described the data, at least at higher $Q^2$, ergo the data (maximum $Q^2$=5.5 GeV$^2$) reached the hard/soft factorization regime. With the finer binning at lower $-t$ afforded by the KaonLT data and access to the new YCK model, we believe there is insufficient evidence to support this conclusion.
We suggest the extraction of GPDs from these data should be delayed until a model-independent test can be performed, for example a $Q^{-n}$ scaling study at fixed $x_B$ of Rosenbluth separated cross-sections. A previous scaling study of pion electroproduction up to $Q^2$=3.9 GeV$^2$ from the F$\pi$-2 Collaboration was inconclusive \cite{Horn_2008_2}, but analysis is ongoing with data from the KaonLT (maximum $Q^2$=5.5 GeV$^2$ \cite{Horn_2008}) and PionLT (maximum $Q^2$=8.5 GeV$^2$ \cite{Gaskell_2009}) experiments.

In summary, the observable $A_{LU}$ and the cross section ratio $\sigma_{LT'}/\sigma_0$ of the $p(\vec{e},e'\pi^+)n$ reaction have been measured at Hall C of Jefferson Lab over a wide range of kinematics. The dependence of $\sigma_{LT'}/\sigma_0$ on $t$ at fixed $(Q^2,~x_B$), and the dependence on $Q^2$ at two fixed $(t,~x_B)$ was explored and compared to theoretical calculations. In all cases, the Regge-based models outperformed the GPD-based GK model, with the best agreement being with YCK1, a Regge model in which the nucleon EMFFs are parametrized with GPDs. This lends support to the hypothesis that the hard-soft factorization scheme is not yet applicable to this reaction in these kinematics.
Rosenbluth separation and factorization scaling studies of the $p(e,e'\pi^+)n$ reaction forthcoming from Hall C will be able to provide a model-independent check of this hypothesis. Additional KaonLT results will include measurements of $\sigma_{LT'}/\sigma_0$ in $p(e,e'\pi^+)\Delta^0$ and $u$-channel meson production, and Rosenbluth separated cross-sections for $p(e,e'K^+)\Lambda/\Sigma$.

\section*{Acknowledgements}
We thank the staff of the Accelerator and the Physics Divisions at Jefferson Lab for their excellent efforts during the experimental data taking.  We also thank Stefan Diehl for providing the $x_B$, $Q^2$ standard deviations of both CLAS data sets. 
This material is based upon work supported by the U.S. Department of Energy, Office of Science, Office of Nuclear Physics under contract DE-AC05-06OR23177.
This work is supported by the Natural Sciences and Engineering Research Council of Canada (NSERC) SAPIN-2021-00026 and a Canadian Institute of Nuclear Physics graduate fellowship. Additional support is gratefully acknowledged from: the University of Regina, the U.S. National Science Foundation (NSF) grants PHY 2309976,  2012430 and 1714133 at the Catholic University of America, NSF grant PHY 2209199 at Ohio University, UK Science and Technology Facilities Council (STFC) grant ST/W004852/1 at the University of York, and National Research Foundation of Korea (NRF) grant NRF-2022 R1A2B5B01002307.

\begin{table*}[t]
    \centering
    \begin{tabular}{c c c c c c c c c c}
    \hline 
    \hline 
        $\langle -t \rangle$ & $-t$ range & $\langle -t_{min} \rangle$  & $\langle Q^2 \rangle$ & $\langle x_B \rangle$ & $\langle \epsilon \rangle$ & $\sigma_{LT'}/\sigma_0$ & $\delta_{\textrm{stat}}$ & $\delta_{\textrm{sys}}^{\downarrow}$ & $\delta_{\textrm{sys}}^{\uparrow}$ \\
       
       [GeV${}^2$] & [GeV${}^2$] & [GeV${}^2$] &  [GeV${}^2$] &  &  &  &  & & \\
        \hline 
     \multicolumn{10}{c}{\rule{0pt}{3ex} (a) Q$^2$=3.0 GeV$^2$, x$_B$=0.40} \\
        \rule{0pt}{3ex} 0.150 & 0.09 -- 0.20 & 0.141 & 2.50 $\pm$ 0.10 & 0.336 $\pm$ 0.017 & 0.890 $\pm$ 0.005 & 0.0852 & 0.0211 & 0.0054 & 0.0118 \\
        0.227 & 0.20 -- 0.25 & 0.168 & 2.67 $\pm$ 0.12 & 0.362 $\pm$ 0.016 & 0.891 $\pm$ 0.006 & 0.0764 & 0.0193 & 0.0075 & 0.0063 \\
        0.275 & 0.25 -- 0.30 & 0.186 & 2.80 $\pm$ 0.15 & 0.378 $\pm$ 0.019 & 0.888 $\pm$ 0.007 & 0.0570 & 0.0155 & 0.0028 & 0.0356 \\
        0.323 & 0.30 -- 0.35 & 0.205 & 2.92 $\pm$ 0.17 & 0.393 $\pm$ 0.021 & 0.886 $\pm$ 0.008 &  0.1133 & 0.0180 & 0.0057 & 0.0060 \\
        0.376 & 0.35 -- 0.40 & 0.223 & 3.02 $\pm$ 0.17 & 0.406 $\pm$ 0.023 & 0.884 $\pm$ 0.009 & 0.1263 & 0.0233 & 0.0115 & 0.0103 \\
        0.447 & 0.40 -- 0.50 & 0.248 & 3.13 $\pm$ 0.17 & 0.424 $\pm$ 0.025 & 0.884 $\pm$ 0.009 & 0.1259 & 0.0167 & 0.0071 & 0.0162 \\
        0.621 & 0.50 -- 0.80 & 0.303 & 3.34 $\pm$ 0.17 & 0.459 $\pm$ 0.029 & 0.885 $\pm$ 0.009 & 0.1479 & 0.0140 & 0.0138 & 0.0022 \\

        \multicolumn{10}{c}{\rule{0pt}{3ex} (b) Q$^2$=4.4 GeV$^2$, x$_B$=0.40} \\
        \rule{0pt}{3ex} 0.154 & 0.09 -- 0.20 & 0.167 & 3.90 $\pm$ 0.16 & 0.358 $\pm$ 0.017 & 0.709 $\pm$ 0.010 & 0.0382 & 0.0107 & 0.0281 & 0.0281 \\
        0.242 & 0.21 -- 0.28 & 0.190 & 4.12 $\pm$ 0.17 & 0.377 $\pm$ 0.015 & 0.724 $\pm$ 0.009 &  0.1012 & 0.0156 & 0.0084 & 0.0026 \\
        0.316 & 0.28 -- 0.35 & 0.208 & 4.30 $\pm$ 0.20 & 0.390 $\pm$ 0.019 & 0.719 $\pm$ 0.011 & 0.1323 & 0.0143 & 0.0053 & 0.0082 \\
        0.398 & 0.35 -- 0.45 & 0.224 & 4.45 $\pm$ 0.22 & 0.402 $\pm$ 0.022 & 0.719 $\pm$ 0.012 & 0.1272 & 0.0153 & 0.0108 & 0.0029 \\
        0.574 & 0.45 -- 0.80 & 0.250 & 4.67 $\pm$ 0.24 & 0.420 $\pm$ 0.026 & 0.722 $\pm$ 0.011 & 0.1356 & 0.0134 & 0.0118 & 0.0032 \\

        \multicolumn{10}{c}{\rule{0pt}{3ex} (c) Q$^2$=5.5 GeV$^2$, x$_B$=0.40} \\
        \rule{0pt}{3ex} 0.183 & 0.11 -- 0.24 & 0.178 & 5.07 $\pm$ 0.18 & 0.365 $\pm$ 0.015 & 0.520 $\pm$ 0.009 &   0.0281 & 0.0254 & 0.033 & 0.0132 \\
        0.271 & 0.24 -- 0.30 & 0.198 & 5.26 $\pm$ 0.19 & 0.381 $\pm$ 0.015 & 0.532 $\pm$ 0.009 & 0.0559 & 0.0191 & 0.0037 & 0.0120 \\
        0.347 & 0.30 -- 0.40 & 0.215 & 5.42 $\pm$ 0.23 & 0.394 $\pm$ 0.018 & 0.530 $\pm$ 0.010 & 0.0880 & 0.0200 & 0.0082 & 0.0030 \\
        0.445 & 0.40 -- 0.50 & 0.230 & 5.55 $\pm$ 0.25 & 0.405 $\pm$ 0.020 & 0.532 $\pm$ 0.011 & 0.1205 & 0.0200 & 0.0065 & 0.0054 \\
        0.616 & 0.50 -- 0.80 & 0.251 & 5.71 $\pm$ 0.25 & 0.418 $\pm$ 0.021 & 0.535 $\pm$ 0.011 & 0.1095 & 0.0187 & 0.0118 & 0.0046 \\

        \multicolumn{10}{c}{\rule{0pt}{3ex} (d) Q$^2$=2.1 GeV$^2$, $x_B$=0.21} \\
        \rule{0pt}{3ex} 0.026 & 0.01 -- 0.04 & 0.034 & 1.74 $\pm$ 0.09 & 0.177 $\pm$ 0.011 & 0.785 $\pm$ 0.009  & 0.0310 & 0.0124 & 0.0037 & 0.0039 \\
        0.055 & 0.04 – 0.07 & 0.041 & 1.87 $\pm$ 0.12 & 0.194 $\pm$ 0.012 & 0.801 $\pm$ 0.008  & 0.0483 & 0.0071 & 0.0021 & 0.0019 \\
        0.085 & 0.07 -- 0.10 & 0.046 & 1.99 $\pm$ 0.15 & 0.204 $\pm$ 0.015 & 0.797 $\pm$ 0.009 & 0.0572 & 0.0075 & 0.0021 & 0.0072 \\
        0.115 & 0.10 - 0.13 & 0.049 & 2.06 $\pm$ 0.17 & 0.210 $\pm$ 0.018 & 0.794 $\pm$ 0.010 & 0.0803 & 0.0073 & 0.0042 & 0.0021 \\
        0.145 & 0.13 -- 0.16 & 0.051 & 2.11 $\pm$ 0.18 & 0.215 $\pm$ 0.019 & 0.792 $\pm$ 0.011 & 0.0929 & 0.0075 & 0.0100 & 0.0044 \\
        0.180 & 0.16 -- 0.20 & 0.054 & 2.16 $\pm$ 0.18 & 0.219 $\pm$ 0.019 & 0.791 $\pm$ 0.012 & 0.1084 & 0.0079 & 0.0104 & 0.0019 \\
        0.244 & 0.20 -- 0.30 & 0.057 & 2.25 $\pm$ 0.18 & 0.225 $\pm$ 0.020 & 0.790 $\pm$ 0.012 & 0.1055 & 0.0071 & 0.0080 & 0.0015 \\
        0.373 & 0.30 -- 0.50 & 0.064 & 2.38 $\pm$ 0.17 & 0.237 $\pm$ 0.020 & 0.789 $\pm$ 0.012 & 0.1079 & 0.0103 & 0.0078 & 0.0036 \\

        \multicolumn{10}{c}{\rule{0pt}{3ex} (e) Q$^2$=3.0 GeV$^2$, x$_B$=0.25} \\
        \rule{0pt}{3ex} 0.046 & 0.02 -- 0.07 & 0.055 & 2.60 $\pm$ 0.11 & 0.221 $\pm$ 0.011 & 0.661 $\pm$ 0.010  & 0.0324 & 0.0145 & 0.0026 & 0.0037 \\
        0.085 & 0.07 -- 0.10 & 0.062 & 2.77 $\pm$ 0.14 & 0.234 $\pm$ 0.012 & 0.677 $\pm$ 0.009 & 0.0626 & 0.0100 & 0.0034 & 0.0018 \\
        0.115 & 0.10 - 0.13 & 0.068 & 2.88 $\pm$ 0.17 & 0.244 $\pm$ 0.014 & 0.678 $\pm$ 0.009 & 0.0653 & 0.0100 & 0.0034 & 0.0047 \\
        0.145 & 0.13 -- 0.16 & 0.071 & 2.95 $\pm$ 0.18 & 0.249 $\pm$ 0.016 & 0.674 $\pm$ 0.010  & 0.0822 & 0.0109 & 0.0045 & 0.0048 \\
        0.179 & 0.16 -- 0.20 & 0.073 & 2.99 $\pm$ 0.20 & 0.252 $\pm$ 0.017 & 0.671 $\pm$ 0.012 & 0.1098 & 0.0092 & 0.0075 & 0.0034 \\
        0.246 & 0.20 -- 0.30 & 0.077 & 3.07 $\pm$ 0.20 & 0.257 $\pm$ 0.019 & 0.670 $\pm$ 0.012 & 0.1084 & 0.0075 & 0.0061 & 0.0022 \\
        0.379 & 0.30 -- 0.50 & 0.082 & 3.21 $\pm$ 0.19 & 0.265 $\pm$ 0.018 & 0.671 $\pm$ 0.012 & 0.1123 & 0.0089 & 0.0071 & 0.0044 \\
        \hline 
        \hline 
    \end{tabular}
    \caption{Summary of all KaonLT measurements of $\sigma_{LT'}/\sigma_0$. The data is divided by experimental setting, each of which is labeled by a central $Q^2$ and $x_B$ and corresponds to one panel of Fig.~\ref{fig:results}. Quantities enclosed in angled brackets denote the mean value of that variable in each $t$-bin. For Mandelstam $t$, the bin borders are provided as a range, whereas for other kinematics the standard deviation of the value is given. The column $\langle t_{min} \rangle$ provides $-t_{min}$ as calculated from $\langle Q^2 \rangle$ and $\langle x_B \rangle$. Since each event has its own $Q^2$ and $x_B$, each event also has its own $-t_{min}$, so the lowest border of the first $t$-bin can be lower than $\langle -t_{min} \rangle$. The variation of $\langle Q^2 \rangle$ and $\langle x_B \rangle$ within a setting is statistically significant, whereas the variation of $\langle \epsilon \rangle$ is not.
    The columns $\delta_{\textrm{stat}}$, $\delta_{\textrm{sys}}^\downarrow$ and $\delta_{\textrm{sys}}^\uparrow$ refer to the uncertainties of $\sigma_{LT'}/\sigma_0$.
    The systematic error is asymmetric, with $\delta_{\textrm{sys}}^\downarrow$ ($\delta_{\textrm{sys}}^\uparrow$) denoting the lower (upper) bounds. The total error in each direction is the sum of $\delta_{\textrm{stat}}$ and $\delta_{\textrm{sys}}^\downarrow$ ($\delta_{\textrm{sys}}^\uparrow$) in quadrature.}
    \label{tab:test}
\end{table*}

\appendix 
\section{Additional Figure on Kinematic Coverage}
 \begin{figure}[hb!]
    \centering
    \includegraphics[width=0.99\linewidth]{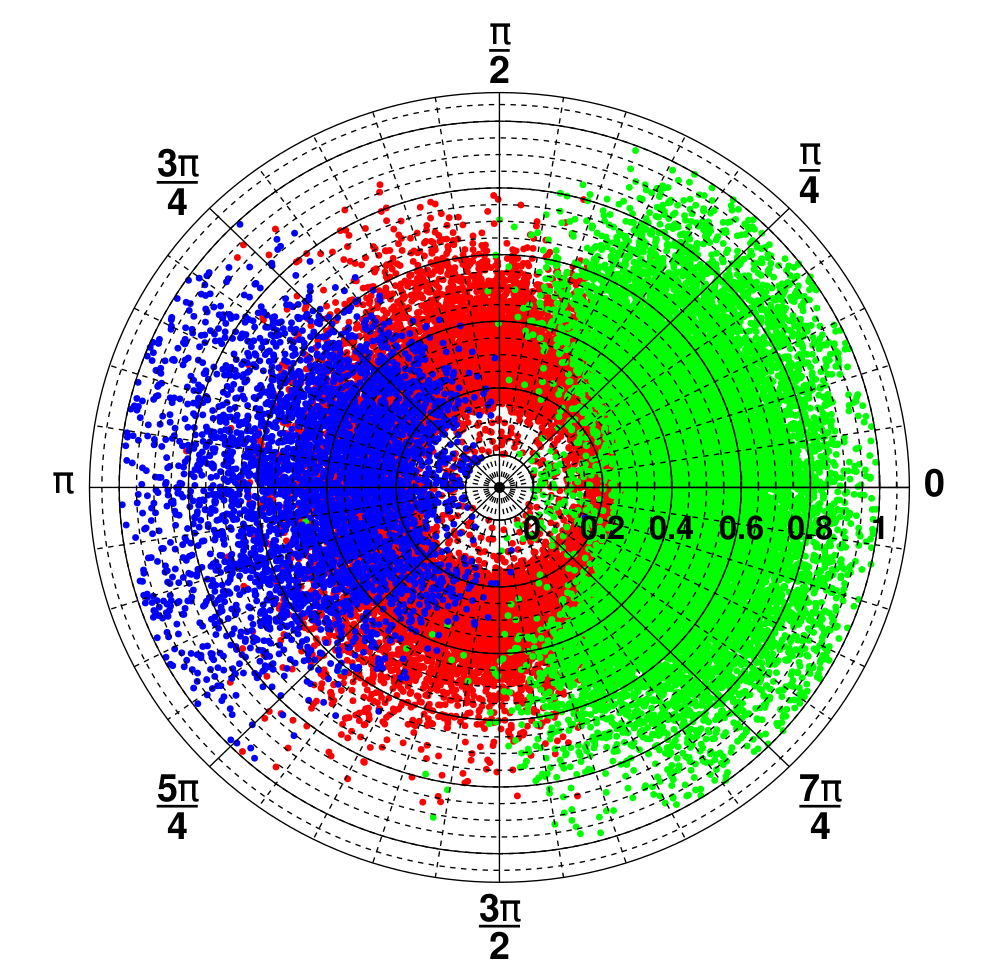}
    \caption{Representative ``$t-\phi$" plot showing the coverage of each SHMS setting. The radial axis is the Mandelstam variable $-t$, and the polar axis is the scattering plane angle $\phi$. Each dot is one event, with the colours corresponding the the settings of the SHMS: blue (left, $-3^\circ$ from $\vec{q}$), red (center, along the direction of $\vec{q}$), green (right, $+3^\circ$ from $\vec{q}$). The setting used to generate this plot is ($Q^2$,$x_B$)=(5.5 GeV$^2$,0.40). The $t$-bins for this setting are rings in azimuth from 0 to $2\pi$ at the values of $-t$ given in Table 1. The $t-\phi$ plots for the other settings are similar.}
    \label{fig:tphi}
\end{figure}

\section{Quantitative Analysis of Models}

In this section, we present the $\chi^2/$NDF for each theoretical model for each setting (Table~\ref{tab:chisq}). The data point at the lowest $-t$ for each setting is excluded from this calculation, as it falls below the minimum $-t$ at the mean kinematics where the model is evaluated. 
The standard formula $\chi^2 = \sum_{i=1}^n (y_i - f(t_i))^2\delta_i^{-2}$ is used, where $y_i$ are the experimental values of $\sigma_{LT'}/\sigma_0$ at $t_i$, $f(t_i)$ are the model predictions for $\sigma_{LT'}/\sigma_0$ at $t_i$, and $\delta_i$ are the experimental uncertainties at each data point. As the models do not include parameters fixed to data, the NDF is simply the number of data points used in the calculation of $\chi^2$.

It is worth noting that the $\chi^2/$NDF of the VR model may be artificially low due to the exact kinematics of this experiment. The VR model fits much better at lowest $-t$, where most of the data in this experiment is taken, and worse at higher $-t$, where there is only one $t$-bin. The $\chi^2/$NDF would be much higher if we included higher $-t$ data, for example the CLAS12 data shown in Fig.5(e).

\begin{table}[h!]
    \centering
    \begin{tabular}{c|c|c|c|c|c}
        \hline 
        \hline 
        Setting & \multicolumn{5}{c}{$\chi^2/\mathrm{NDF}$}\\
        (Q$^2$ [GeV$^2$], x$_B$) & VR & YCK1 & YCK2 & GK1 & GK2 \\
        \hline
        (3.0, 0.4) & 1.4 & 0.5 & 9.9 & 20.0 & 2.8 \\
        (4.4, 0.4) & 8.0 & 2.6 & 11.6 & 44.0 & 7.0 \\
        (5.5, 0.4) & 20.1 & 6.2 & 8.4 & 89.7 & 21.2 \\
        (2.1, 0.21) & 8.9 & 9.2 & 27.3 & 173.7 & 94.8 \\
        (3.0, 0.25) & 10.9 & 4.3 & 27.3 & 168.4 & 80.2 \\
        Average & 9.9 & 4.6 & 16.9 & 99.2 & 41.2 \\
        \hline 
        \hline 
    \end{tabular}
    \caption{The reduced $\chi^2$ comparing experimental $\sigma_{LT'}/\sigma_0$ from the KaonLT experiment to model calculations at each setting. The models are described in the text and exact values of data for each setting are provided above.}
    \label{tab:chisq}
\end{table}

\section{Asymmetry Plots}

The following figures contain the asymmetry $A_{LU}$ as a function of $\phi$ in each $-t$ bin. The asymmetry was experimentally calculated as follows, with a statistical uncertainty calculated using the general formula for error propagation and counting statistics $\delta(Y^{\pm})=\sqrt{Y^{\pm}}$
\begin{equation}
    A_{LU} = \frac{1}{P}\bigg(\frac{Y^+-Y^-}{Y^++Y^-}\bigg), \hspace{1cm} \delta_{\textrm{stat}} = \frac{2}{P}\sqrt{\frac{Y^+Y^-}{(Y^++Y^-)^3}}
    \label{eqn:bsa}
\end{equation}

$\sigma_{LT'}/\sigma_0$ is extracted from $A_{LU}$ according to the equation
\begin{equation}
    A_{LU} = \frac{\sqrt{2\epsilon(1-\epsilon)} \frac{\sigma_{LT'}}{\sigma_0}\sin\phi}{1+ \sqrt{2\epsilon(1+\epsilon)}\frac{\sigma_{LT}}{\sigma_0}\cos\phi + \epsilon\frac{\sigma_{TT}}{\sigma_0}\cos2\phi} 
    \label{full}
\end{equation}
where  $\sigma_0=\sigma_T+\epsilon\sigma_L$ is the unpolarized cross section,
$\epsilon$ is the ratio of longitudinal and transverse virtual photon polarization, $\sigma_{LT}$, $\sigma_{TT}$, $\sigma_{LT'}$ are interference cross sections, and $\phi$ is the scattering plane angle. $A_{LU}$ is sometimes approximated as
\begin{equation}
    A_{LU} = \sqrt{2\epsilon(1-\epsilon)} \frac{\sigma_{LT'}}{\sigma_0}\sin\phi
    \label{approx}
\end{equation}
where the terms $\sigma_{LT}/\sigma_0$ and $\sigma_{TT}/\sigma_0$ are considered to be negligible. In this work, $\sigma_{LT'}/\sigma_0$ is extracted using Eqn. \ref{full}, and the value from Eqn. \ref{approx} is used to determine a systematic uncertainty.

The exact values of $A_{LU}$ as a function of $\phi$ in each bin are available on Mendeley data:
\vspace{3mm}

\noindent 
A.C. Postuma for the KaonLT Collaboration (2025), \textit{KaonLT BSA Analysis of p(e,e'pi+)n,} Mendeley Data, V1

\noindent
\href{https://data.mendeley.com/datasets/m47c5t4ykg/1}{doi:10.17632/m47c5t4ykg.1}

\begin{figure*}[h!]
    \centering\includegraphics[width=0.85\linewidth]{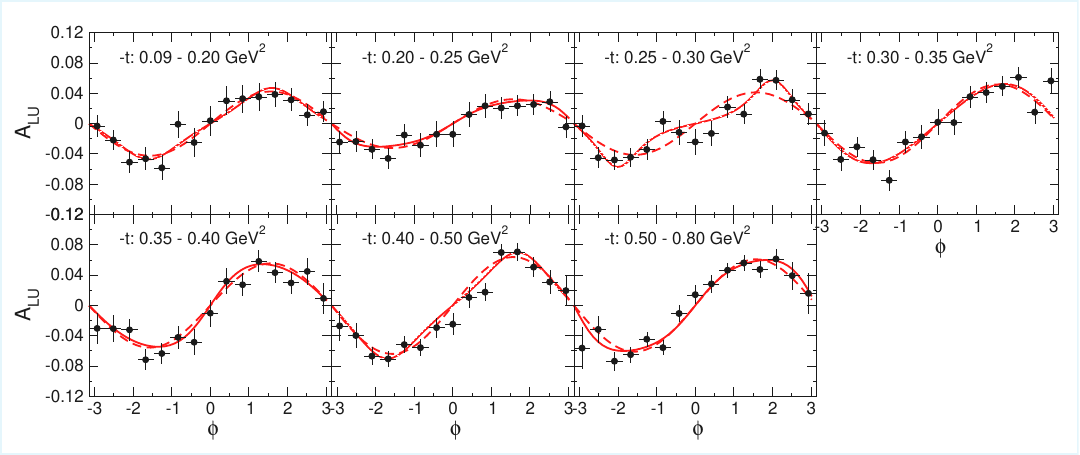}
    \caption{Binned asymmetry for setting 1, at central values of Q$^2$=3.0, x$_B$=0.4. The solid line shows the full fit according to Eqn. \ref{full}, and the dashed line shows the approximated fit from Eqn. \ref{approx}. Uncertainties are statistical only.}
\end{figure*}

\begin{figure*}[h!]
    \centering
    \includegraphics[width=0.85\linewidth]{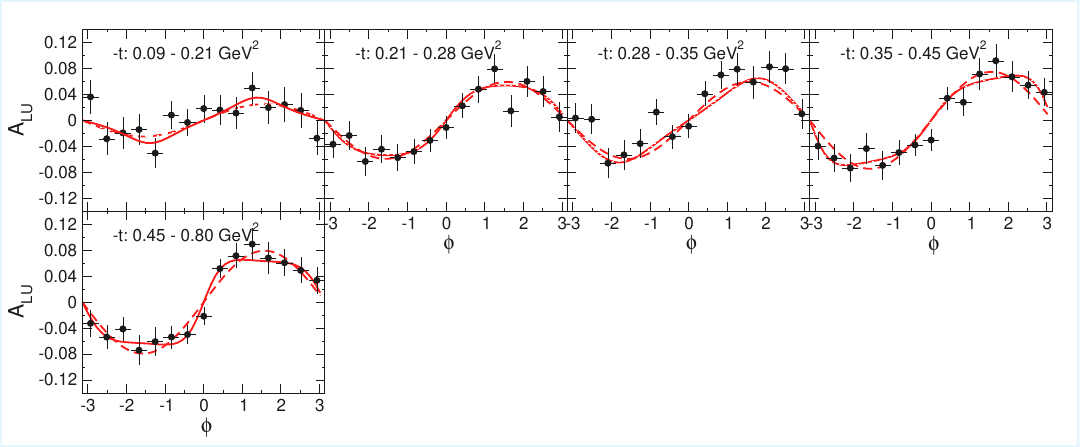}
    \caption{Binned asymmetry for setting 2, at central values of Q$^2$=4.4, x$_B$=0.4. The solid line shows the full fit according to Eqn. \ref{full}, and the dashed line shows the approximated fit from Eqn. \ref{approx}. Uncertainties are statistical only.}
\end{figure*}

\begin{figure*}[h!]
    \centering\includegraphics[width=0.85\linewidth]{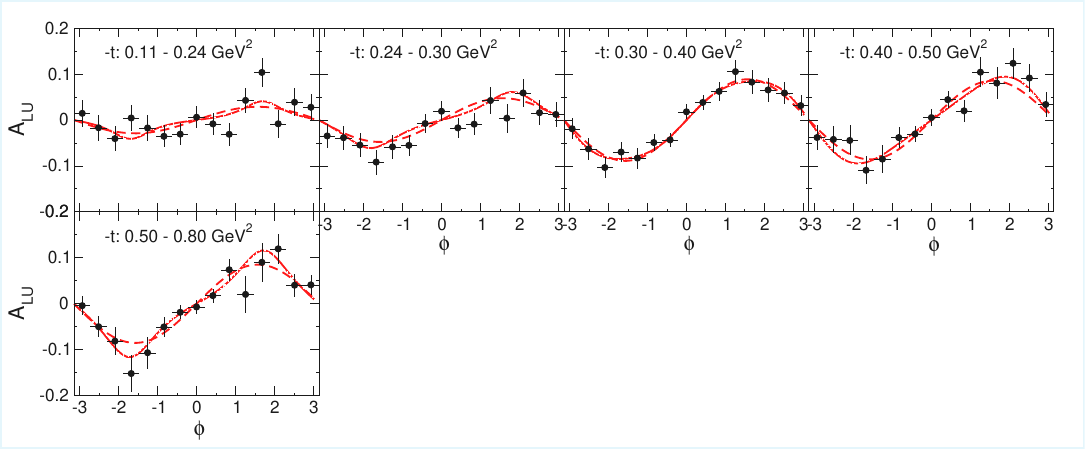}
    \caption{Binned asymmetry for setting 3, at central values of Q$^2$=5.5, x$_B$=0.4. The solid line shows the full fit according to Eqn. \ref{full}, and the dashed line shows the approximated fit from Eqn. \ref{approx}. Uncertainties are statistical only.}
\end{figure*}

\begin{figure*}[h!]
    \centering\includegraphics[width=0.85\linewidth]{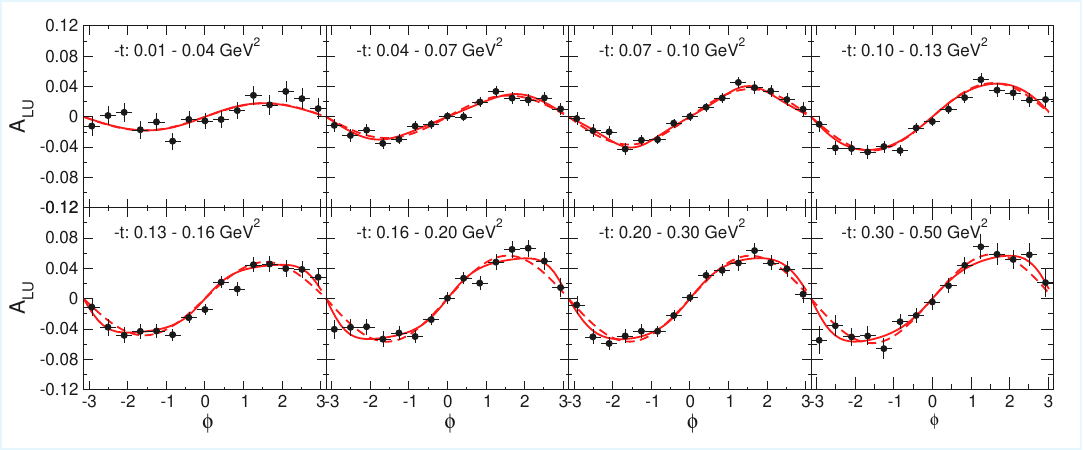}
    \caption{Binned asymmetry for setting 4, at central values of Q$^2$=2.1, x$_B$=0.21. The solid line shows the full fit according to Eqn. \ref{full}, and the dashed line shows the approximated fit from Eqn. \ref{approx}. Uncertainties are statistical only.}
\end{figure*}

\begin{figure*}[h!]
    \centering\includegraphics[width=0.85\linewidth]{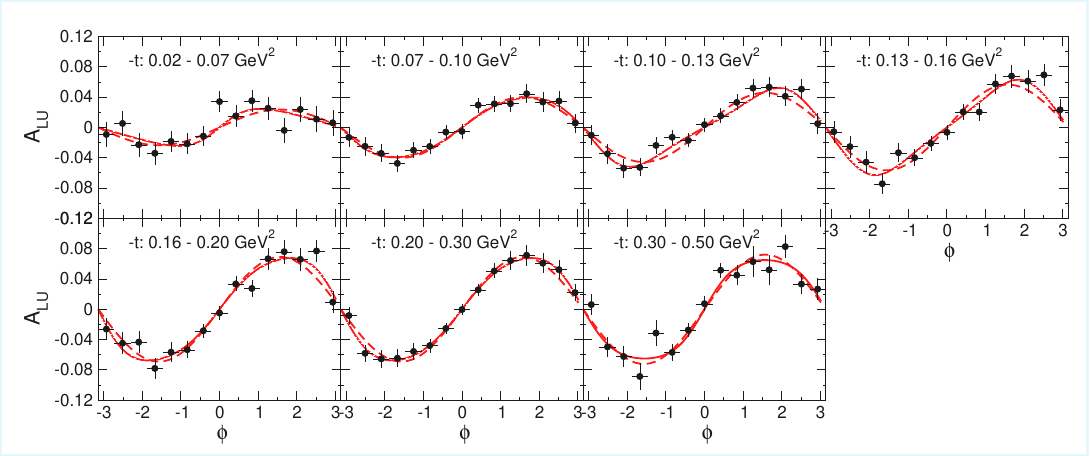}
    \caption{Binned asymmetry for setting 5, at central values of Q$^2$=3.0, x$_B$=0.25. The solid line shows the full fit according to Eqn. \ref{full}, and the dashed line shows the approximated fit from Eqn. \ref{approx}. Uncertainties are statistical only.}
\end{figure*}

\end{document}